\documentclass[aps,pra,twocolumn,preprintnumbers,showpacs,superscriptaddress,amsmath,amssymb]{revtex4}
\usepackage{dcolumn}
\usepackage{bm}
\usepackage{amssymb}
\usepackage[german, english]{babel}
\usepackage{graphicx}
\usepackage{color}
\usepackage{textcomp}
\usepackage[letterpaper,total={7in,9.5in},top=0.75in,left=0.75in]{geometry}
\usepackage[utf8]{inputenc}

\usepackage{array}

\begin{document}

\title{Converting single photons from an InAs/GaAs quantum dot into the ultraviolet: preservation of second-order correlations}

\author{Anica Hamer}
\thanks{These authors contributed equally.}
\affiliation{Physikalisches Institut, Rheinische Friedrich-Wilhelms-Universität, 53115 Bonn, Germany}
\author{David Fricker}
\thanks{These authors contributed equally.}
\affiliation{Peter Grünberg Institute 9, Forschungszentrum Jülich GmbH, Jülich, Germany}
\affiliation{Department of Physics, RWTH Aachen University, Aachen, Germany}
\author{Marcel Hohn}
\affiliation{Physikalisches Institut, Rheinische Friedrich-Wilhelms-Universität, 53115 Bonn, Germany}
\author{Paola Atkinson}
\affiliation{Institut des Nano Sciences de Paris, CNRS UMR 7588, Sorbonne Universite, France}
\author{Mihail Lepsa}
\affiliation{Peter Grünberg Institute 9, Forschungszentrum Jülich GmbH, Jülich, Germany}
\author{Stefan Linden}
\affiliation{Physikalisches Institut, Rheinische Friedrich-Wilhelms-Universität, 53115 Bonn, Germany}
\author{Frank Vewinger}
\affiliation{Institut für Angewandte Physik, Rheinische Friedrich-Wilhelms-Universität Bonn, Bonn, Germany}
\author{Beata Kardynal}
\affiliation{Peter Grünberg Institute 9, Forschungszentrum Jülich GmbH, Jülich, Germany}
\affiliation{Department of Physics, RWTH Aachen University, Aachen, Germany}
\author{Simon Stellmer}
\email{stellmer@uni-bonn.de}
\affiliation{Physikalisches Institut, Rheinische Friedrich-Wilhelms-Universität, 53115 Bonn, Germany}

\date{\today}


\begin{abstract}

Wavelength conversion at the single-photon level is required to forge a quantum network from distinct quantum devices. Such devices include solid-state emitters of single or entangled photons, as well as network nodes based on atoms or ions. Here we demonstrate the conversion of single photons emitted from a III-V semiconductor quantum dot at 853\,nm via sum frequency conversion to the wavelength of the strong transition of Yb$^+$ ions at 370\,nm. We measure the second-order correlation function of both the unconverted and of the converted photon and show that the single-photon character of the quantum dot emission is preserved during the conversion process.

\end{abstract}


\maketitle

\section{Introduction}
Quantum networks, as envisioned for quantum computation and quantum communication applications \cite{Kimble2008,Ladd2010}, are often based on a hybrid architecture. Such a layout may include solid-state emitters, network nodes based on single or few atoms or ions, and photons as so-called flying qubits \cite{Meyer2015,Bock2018,Schupp2021}. This approach requires an efficient and entanglement-preserving exchange of photons between the individual components. Such an interconnect involves frequency conversion of the photon \cite{Jayakumar2014,Matutano2016,Rutz2017,Krutyanskiy2017,Weber2019,Leent2020}. Over the past years, a number of studies have demonstrated that quantum frequency conversion can bridge the gap between the visible or infrared (IR) emission of solid-state devices and the telecom band \cite{Tanzilli2005,Matutano2016,Krutyanskiy2017,Weber2019,Leent2020}. Conversion to the ultraviolet (UV) spectral region, where strong transitions in atoms and ions can be found, remains a challenge; see e.g. Ref.~\cite{Rutz2017}.

Here, we report on experimental progress towards the transfer of a spin state between an exciton in a self-assembled InAs/GaAs quantum dot (QD) and an ytterbium (Yb) ion. Such state transfer will facilitate a quantum link between spin qubits in these systems. The Yb${}^+$ ion can be coupled to an optical cavity to form a network node \cite{Kobel2021}. While self-assembled quantum dots have been demonstrated as an excellent interface between photons and spins \cite{Gao2012}, gated quantum dots feature longer spin-state lifetime and can be scaled up into quantum processors. Self-assembled quantum dots with wavelength of emission between 850\,nm and 870\,nm have been proposed as an interface for the transfer of a photonic state, via the spin of a confined excitonic state, to the spin state in a gate-confined quantum dot \cite{Joecker2019}. Subsequently, the spin transfer concept requires quantum frequency conversion between the InAs/GaAs quantum dot emission wavelength around 853\,nm and the principal resonance in the Yb$^+$ ion near 370\,nm. Conversion between IR and UV wavelengths is challenging for two reasons: the large wavelength gap and crystal absorption in the UV. Infrared to UV frequency conversion of single photons was reported only recently by the group of C. Silberhorn \cite{Rutz2017}.

In this Letter, we present a 853\,nm $\rightarrow$ 370\,nm frequency conversion module with an external efficiency of about 0.35\% at a background photon rate of 700\,s$^{-1}$. The obtained signal-to-background allows us to measure, for the first time, the correlation function of single UV photons converted from the emission of an InAs/GaAs quantum dot. This work constitutes a step towards the coherent coupling of solid-state emitters and ions, and brings us closer to the realization of a quantum network utilizing different systems.

\section{Experimental setup}

\begin{figure}[htpb]
	\centering
	\includegraphics[width=\linewidth]{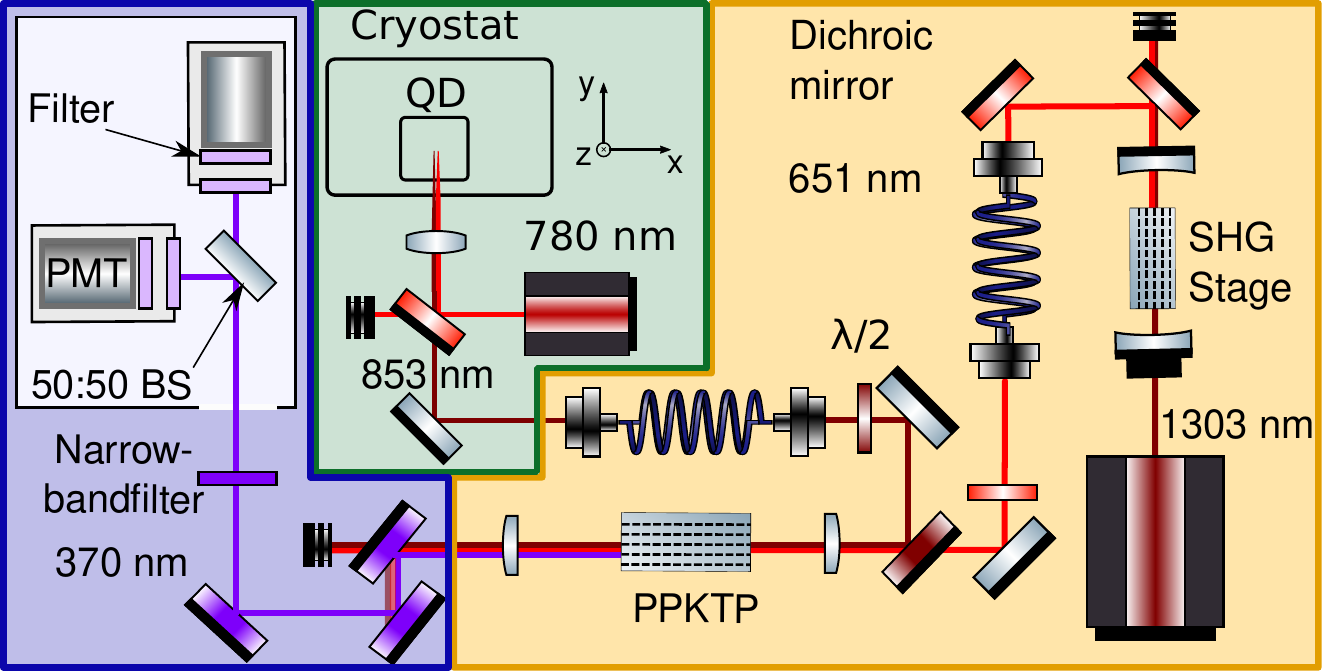}	
	\caption{Schematic overview of the experimental setup, subdivided into the single photon source (green), the frequency conversion module with the excitation laser (yellow), and the detection module for converted photons (blue).}
	\label{fig:QD_sketch}
\end{figure}

The experimental setup is subdivided into the single-photon source, the frequency conversion module with the required pump laser, and the detection module; see Fig.~\ref{fig:QD_sketch}. The configuration and characteristics of the individual components will be described in the following.

\subsection{The single photon source}
The GaAs-based heterostructures containing InAs quantum dots used in this experiment were grown on a (100)-GaAs substrate by molecular beam epitaxy. Quantum dots were formed using a droplet epitaxy mode \cite{Skiba2017}, to benefit from the suppression of the wetting layer and the small natural fine structure splitting of such quantum dots \cite{Skiba2017,Anderson2020}. The wafer contains a low-density ensemble of quantum dots emitting in the wavelength range of 850\,nm to 870\,nm, as needed for the potential coupling to a gated quantum dot \cite{Joecker2019}. The wetting layer emission is near 839\,nm. The layer of quantum dots is located in a weakly coupling cavity formed between a GaAs/AlAs distributed Bragg mirror (DBR) and the GaAs/air surface. The quantum dots are located 118\,nm above the DBR and capped with 105\,nm GaAs.

For the experiment, the sample is mounted on a xyz-positioning stage and cooled down to 1.6\,K in a closed-cycle cryostat (Attodry 2100). It is excited with 780\,nm CW radiation, reflected from a dichroic mirror towards an achromatic lens with a numerical aperture of 0.81. This lens focuses the excitation beam on the sample and collects the photoluminescence signal, which is then coupled into a polarization-maintaining (PM) single-mode fiber. The total extraction and fiber coupling efficiency of emission from the biexciton recombination is estimated as 0.9\,\%.

For the photoluminescence measurements, the quantum dot emission is guided to a Czerny-Turner spectrometer equipped with a CCD (charge coupled device) camera. For the correlation measurements, the sample light is guided to a mechanically adjustable transmission grating and coupled into a 50:50 single-mode fiber splitter. Each output fiber is connected to a superconducting nanowire single-photon detector (SNSPD, Single Quantum Eos CS).

\begin{figure}[t]
	\centering
	\includegraphics[width=\linewidth]{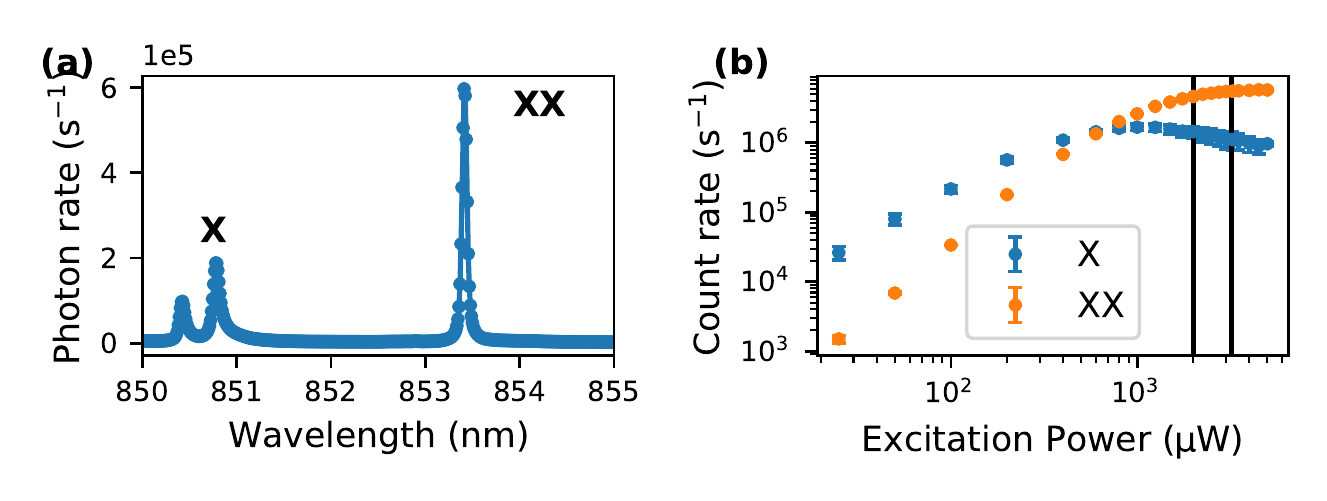}
	\caption{Quantum dot emission resolved by a spectrometer: (a) photoluminiscence spectrum, showing the exciton (X) and biexciton (XX) peaks, and (b) emission rate in dependence of the relative excitation power. Experiments are performed at 2\,mW and 3.2\,mW, indicated by black lines.}
	\label{fig:power_overview1}
\end{figure}

The spectrum of the quantum dot, selected for this experiment, is shown in Fig.~\ref{fig:power_overview1}(a). It consists of exciton (X) and biexciton (XX) lines, at 850.8\,nm and 853.4\,nm, respectively. Figure \ref{fig:power_overview1}(b) shows the dependence of the exciton and biexciton line intensities on excitation power. The maximum intensity of the biexciton emission under CW excitation is higher than that of the exciton. To maximize the photon rate, we use the biexciton emission signal in this work.

Photons emitted by the quantum dot are linearly polarized along two orthogonal directions with equal probability. Both polarizations are delivered to the frequency conversion module via the polarization-maintaining optical fiber, but only one polarization is converted in the crystal. A coupler between these two modules incurs an additional loss of 40\%. For the wavelength conversion experiment, two different values for the excitation power are chosen:~2\,mW (measured before the fiber input), which corresponds to the onset of saturation of the biexciton emission signal, and 3.2\,mW. For these two values, we obtain photon rates of $1.4\times10^6\,{\rm s}^{-1}$ and $1.7\times10^6\,{\rm s}^{-1}$, respectively, in each polarization, before the PPKTP crystal.

\subsection{The frequency conversion module}

We seek to convert single photons emitted by the InAs/GaAs quantum dot, in the range between 853.2\,nm and 853.6\,nm, to the wavelength of the $^2S_{1/2} \rightarrow {}^2 P_{1/2}$ transition in Yb$^+$ ions (369.42\,nm). To this end, we employ sum frequency conversion in periodically poled potassium titanyl phosphate (PPKTP, obtained from AdvR, poling period $2.2\,\mu$m, length 6\,mm) \cite{Rutz2017} with pump light in the range of 651.3\,nm to 651.7\,nm. The pump light is generated by resonant second-harmonic generation (SHG) with a periodically poled lithium niobate (PPLN) crystal from an amplified solid-state laser near 1300\,nm:~this approach combines sufficient wavelength tunability with comparably high output powers in the range of a few 100\,mW.  With the biexciton emission wavelength of 853.42\,nm, the pump wavelength required to match the Yb$^+$ resonance is 651.39\,nm.

Dichroic mirrors and achromatic optics are used to overlap and focus the quantum dot emission and the red pump light into the waveguide structure of the PPKTP crystal. Behind the output of the crystal, the three light fields are collimated and the UV photons are separated from the pump laser and quantum dot emission with dichroic mirrors. The UV light is passed through three bandpass filters:~two filters with FWHM bandwidths of 6\,nm and 10\,nm, respectively, both over 94\% transmission near 370\,nm, and one tunable narrow-band filter with 0.5\,nm bandwidth and 67.8\% transmission.

\begin{figure}[t]
	\centering
	\includegraphics[width=\linewidth]{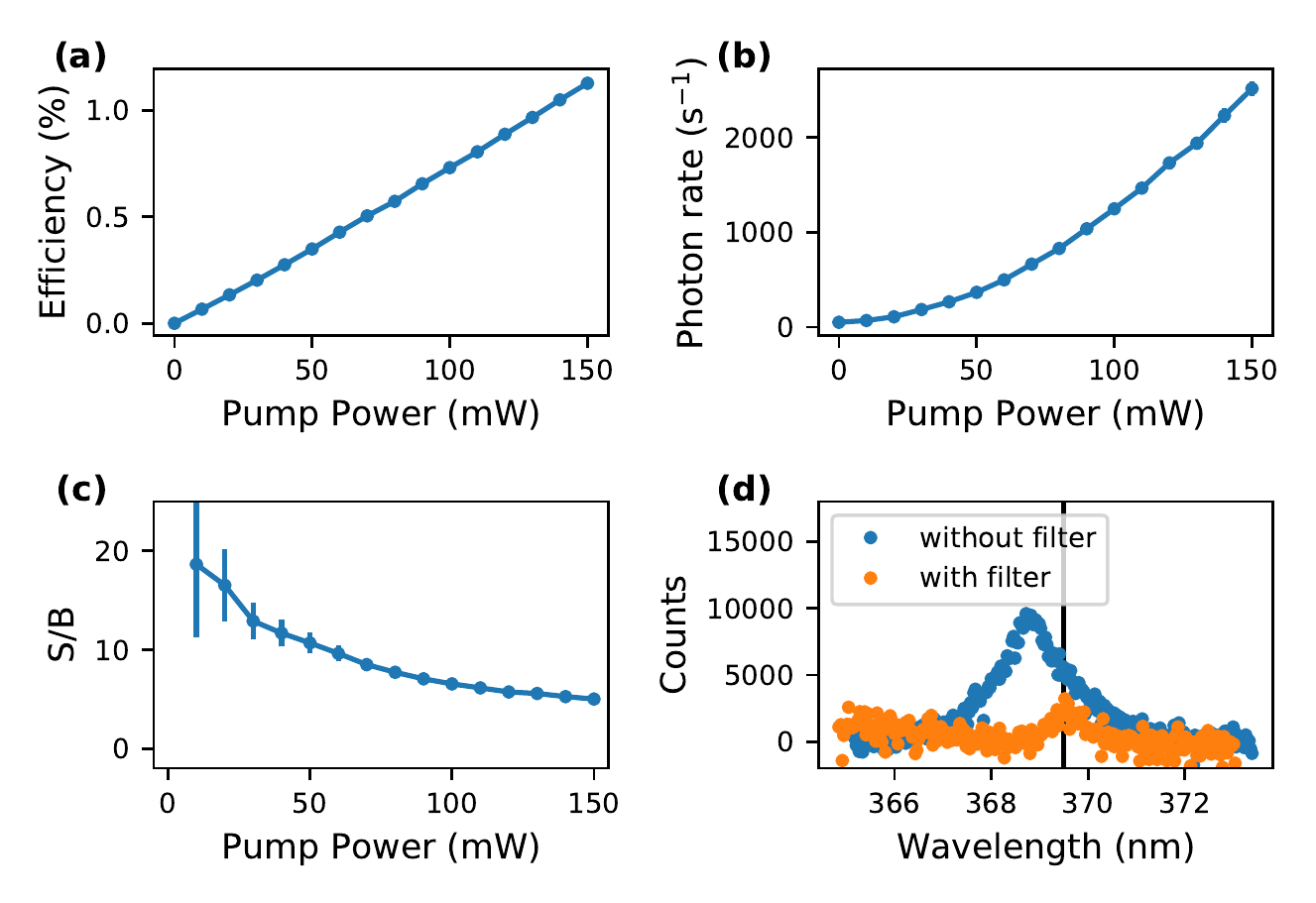}
	\caption{Frequency conversion with a CW auxiliary laser. (a) Efficiency in dependence of pump power, (b) undesired USPDC background in dependence of pump power, (c) signal-to-background ratio, and (d) spectrum of the USPDC background. A narrow-band filter, centered about the target transition wavelength in the Yb$^+$ ion near 369.5\,nm (vertical black line), reduces the background by a factor of 30.}
	\label{fig:overview}
\end{figure}

A wavelength tunable laser is set to the same wavelength as the biexcition signal for characterization and optimization of the conversion module. Upon initial setup of the module, we had obtained an external conversion efficiency $\eta$ of about 15\%/W of pump light. This was about 25\% less than calculated from crystal parameters \cite{Kleinman1966} and specified by the manufacturer. We suspect that mechanical stress from the mounting has caused this reduction in efficiency. During preliminary work, the efficiency reduced to 7.15\%/W due to crystal degradation. This value is reduced to 4.3\%/W when including losses on the optics and filters (device efficiency). The conversion efficiency scales approximately linearly with pump power; see Fig.~\ref{fig:overview}(a).

The conversion in PPKTP is highly temperature dependent, where the FWHM of the efficiency curve is 1.0(1)\,K. Similarly, the conversion efficiency depends on the wavelength of the input photons, and we determine the FWHM of the efficiency curve to be 0.2\,nm. As a consequence, exciton emission of the quantum dot, which is 3\,nm away from the biexciton emission, is not converted to the UV.

The pump light generates undesired up-converted spontaneous parametric down conversion (USPDC) light within the phase-matching window of the PPKTP crystal \cite{Rutz2017}. The amplitude of this background scales quadratically with pump power. At a pump power of 150\,mW, about $85\times10^3$\,photons/s are generated. The spectrum is shown in Fig.~\ref{fig:overview}(d):~it is centered at 368.84\,nm with a FWHM of 1.53\,nm. To reduce this background, we insert the tunable narrow-band filter of 0.5\,nm width. This filter reduces the background-induced count rate on the detectors by a factor of 30, at a loss of signal count rate of only 32.2\%; see Fig.~\ref{fig:overview}(b). The pump power is adjusted between 80\,mW and 150\,mW for a compromise between signal count rate and signal-to-background ratio; see Fig.~\ref{fig:overview}(c).

\subsection{The detection modules}

We employ two different detection modules. For direct detection of the 853\,nm photons emitted by the quantum dot, we employ two SNSPDs with a detection efficiency of 80(2)\%, timing jitters of 18\,ps and 23\,ps, respectively, and a dark count rate of 50 counts per second (cps). Background light is removed from the quantum dot photoluminescence by insertion of a transmission grating (> 90\% transmission at 853\,nm) before the single mode fiber.

For detection of the frequency-converted photons near 370\,nm, we utilize photo-multiplier tubes (PMTs, Hamamatsu model H10682). The detection efficiencies are 36\% and 40\%, respectively, the typical timing jitter is 330\,ps, and the dark count rates are 5\,cps and 20\,cps, respectively. The PMTs are protected from residual pump- and straylight by two OD 4 bandpass filters.

In both detection modules, the incoming light is split on a polarization-independent 50:50 beam splitter to form a Hanbury Brown and Twiss interferometer. 

We employ a counting module to measure the second-order correlation function, $g^{(2)}(\tau)$. A wavelength tunable Ti-sapphire laser at 853\,nm (4\,ps pulses, 80\,MHz repetition rate) is used to characterize timing jitters and delays of the detection system of the upconverted UV photons. We determine the instrumental timing jitter as $\sigma = 213(11)\,$ps, largely limited by the PMTs, and the differential delay between the two detectors as $t_0 = -167(12)\,$ps.

\section{Second order correlation measurements}

\subsection{$g^{(2)}$ function of the quantum dot emission}

\begin{figure}[t]
	\centering
	\includegraphics[width=\linewidth]{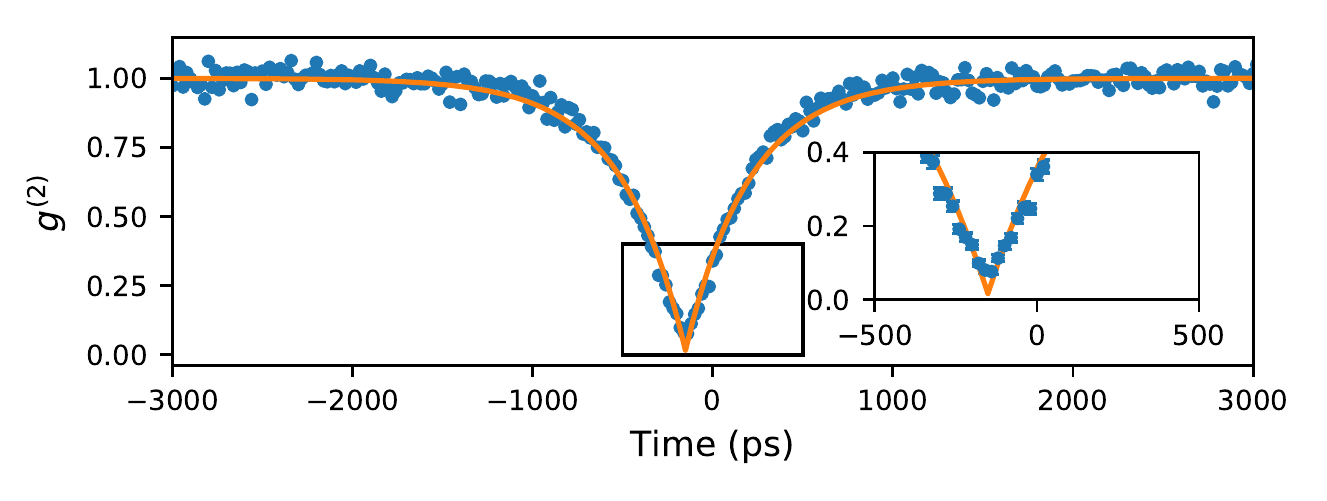}
	\caption{$g^{(2)}$ function of the quantum dot emission. Error bars are within the symbols. The data is fitted by the second order correlation function of an emitter with Lorentzian profile.}
	\label{fig:g2_david}
\end{figure}

We excite the quantum dot with a power of 2\,mW and record the $g^{(2)}$-function of the biexciton emission line, shown in Fig.~\ref{fig:g2_david}. 
The data can be fitted by the second order correlation function of a Lorentzian spectrum,
\begin{equation}
g^{(2)}(\tau) = 1 - a \cdot \exp\left( -\dfrac{|\tau-t_0|}{\tau_0}\right),
\end{equation}
where $\tau_0$ is the lifetime of the biexciton, $a$ is the contrast of the dip, $\tau$ is the time delay, and $t_0$ is the differential electronic delay between the two detectors \cite{Matutano2016}. We obtain a value of $g^{(2)}(t_0) = 0.016(4)$ and a correlation time of $\tau_{0} = 357(3)\,$ps, which corresponds to the biexciton lifetime. The timing jitter of the photodetectors is much shorter and is ignored in the fit.

We also study the dependence of $g^{(2)}(t_0)$ and $\tau_0$ on the excitation power. For increased power, the biexciton lifetime is reduced. For an excitation power of 3.2\,mW, we obtain $g^{(2)}(t_0) = 0.048(7)$ and $\tau_{0} = 264(4)\,$ps. The lifetime reduction is a result of increasingly fast refilling of the quantum dot after photon emission \cite{Michler2000}. At the same time, the contrast of the $g^{(2)}$ dip is reduced, presumably due to the increased background level.

\subsection{$g^{(2)}$ function of the frequency-converted photons} 

Before measuring the $g^{(2)}(\tau)$ function of the frequency-converted photons, we perform extensive tests in which we replace the quantum dot by a CW laser, at the quantum dot's emission wavelength, attenuated to a rate of $1.7 \times 10^6$ photons/s, as expected from the quantum dot. We adjust the pump power to about 100\,mW and choose an integration time of 65\,hours. For a time bin width of 200\,ps, we record on average 124 coincidences per bin. The data follows purely Poissonian statistics. We obtain $g^{(2)}(\tau)=1$, as expected for a coherent light source. Performing the same experiment over 17\,h without the input photons at 853\,nm yields an average of 1.7 coincidences per 200-ps bin. This value is consistent with a USPDC rate of about 2000 photons/s just before the beamsplitter.

We then excite the quantum dot with 2\,mW of power and feed the frequency conversion stage with $1.4\times10^6$\,photons/s from the biexciton emission. The pump power is set to 77(5)\,mW, where the uncertainty accounts for power fluctuations over the duration of the measurement. A correlation measurement of the converted photons is performed for a total of 122\,h and yields on average 85 coincidences per 200-ps bin.

The measured coincidences are normalized to the mean value outside of the range from -3\,ns to +3\,ns. The normalized data is fitted by a convolution of the ideal $g^{(2)}$ function and a Gaussian distribution to account for the timing jitters,
\begin{align}\label{Eq:convolution_fit}
g^{(2)}_\text{exp}(\tau) = \dfrac{1}{\sigma \sqrt{2\pi}}&\int\limits_{-\infty}^\infty  \exp\left(-\dfrac{(\tau-\Tilde{\tau})^2}{2\sigma^2}\right)\nonumber\\
&\text{x}\left( 1 - a \cdot \exp\left( -\dfrac{|\Tilde{\tau}-t_0|}{\tau_0}\right)\right)\text{d}\Tilde{\tau},
\end{align}
where $\tau$ is the delay time, $\tau_0$ is the decay time of the quantum dot, $a$ is the contrast of the dip, and $\sigma$ is the FWHM of the instrumental response \cite{Matutano2016}.

\begin{figure}[t]
	\centering
	\includegraphics[width=\linewidth]{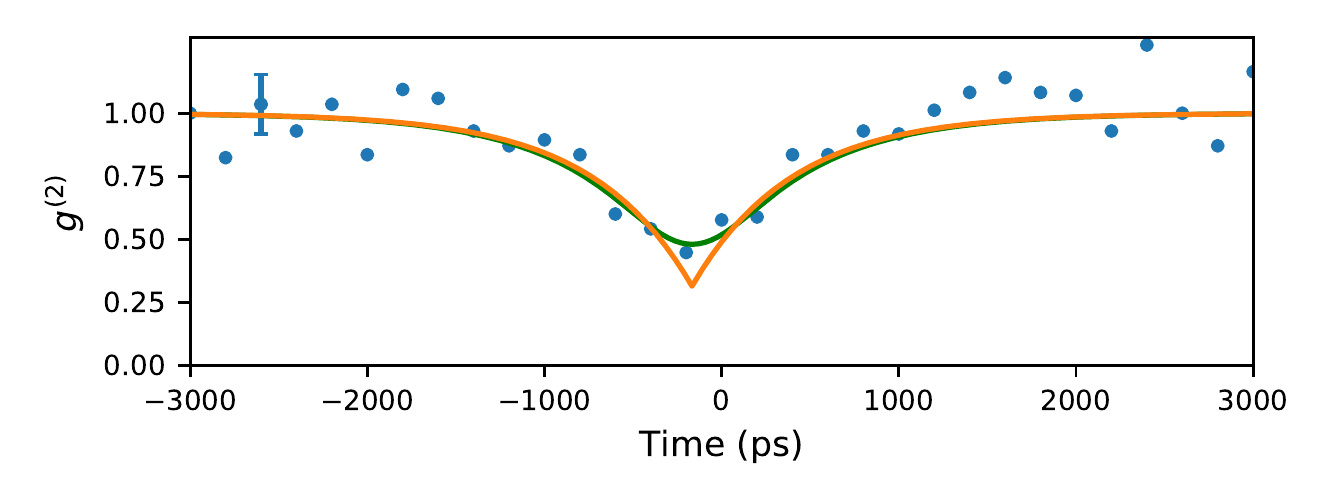}
	\caption{$g^{(2)}$ function of the frequency-converted quantum dot emission. A representative error bar is shown. No background has been subtracted. The data (circles) is fitted by the second order correlation function of an emitter with a Lorentzian profile, convoluted with the Gaussian distribution of the timing jitters (green line). The derived correlation function is represented by an orange line.}
	\label{fig:g2_ani}
\end{figure}

The instrumental shift $t_0$ and jitter $\sigma$ have been determined earlier are used as fixed parameters for the fit. The data are shown in Fig.~\ref{fig:g2_ani}, together with the model from Eq.~(\ref{Eq:convolution_fit}) and the derived $g^{(2)}(\tau)$ function without the instrumental broadening effects. We obtain a correlation time of $\tau_0 = 565(110)\,$ps.

From this analysis, we obtain $g^{(2)}(t_0) = 0.31(8)$. The reduction of contrast in comparison to the direct $g^{(2)}$ measurement is explained by the contribution of the USPDC background. The 85 coincidences recorded on average per bin are comprised of 65(9) signal/signal coincidences, 19(4) signal/background coincidences, and 1.4(4) background/background coincidences. Again, the uncertainties account for an estimated 6\% variation of the pump power during the measurement. When removing the signal/background and background/background coincidences from the dataset, we arrive at $g^{(2)}(t_0) = 0.07(9)$, in agreement with the direct $g^{(2)}$ measurement.

We then perform the same experiment for an increased excitation power of 3.2\,mW, analogous to the direct $g^{(2)}$ measurement in the infrared. With an incident photon rate of $1.7\times10^6$\,photons/s, 98\,h of integration, and 81(5)\,mW of pump light, we find $\tau_0 = 454(115)\,$ps and $g^{(2)}(t_0) = 0.43(9)$. From a total of 106 recorded coincidences on average, 84(10) are signal/signal, 21(4) are signal/background, and 1.5(3) are background/background. Carving out only the signal/signal coincidences, we arrive at $g^{(2)}(t_0) = 0.22(10)$.

\section{Conclusion} 

We have implemented a frequency conversion module to convert single photons emitted by an InAs/GaAs quantum dot near 853\,nm to a wavelength of about 370\,nm. We have shown that the correlation function of the single-photon emitter is preserved throughout the conversion. The measurement became possible only through spectral filtering of the USPDC background generated by the pump light, more advanced filtering will remove this background further. The photon extraction efficiency can be improved by adding a  light-coupling structure, e.g., a solid immersion microlens, on the substrate \cite{Gschrey2015,Chen2018}. Further, the conversion efficiency can be drastically increased from its current value of 0.35\,\% by implementation of a resonant build-up cavity for the pump light \cite{Albota2004}.

As a next step, we will modify the conversion module to convert input light of both polarizations \cite{Bock2018}. This will constitute a quantum frequency conversion stage between two distinct quantum emitters, in our case realized by a quantum dot and an Yb$^+$ ion.

\begin{acknowledgments}
We thank Michael Köhl and all members of the ML4Q Cluster for stimulating discussions, and we thank Christoph Krause and Benjamin Bennemann for technical support in growing the sample. Access to the Helmholtz Nano Facility (HNF) is acknowledged. We acknowledge funding by Deutsche Forschungsgemeinschaft DFG through grant INST 217/978-1 FUGG and through the Cluster of Excellence ML4Q (EXC 2004/1 – 390534769).

Data underlying the results presented in this paper are not publicly available at this time but may be obtained from the authors upon reasonable request.
\end{acknowledgments}

\bibliographystyle{apsrev}
\bibliography{bib}

\end{document}